\documentclass[conference]{IEEEtran}
\IEEEoverridecommandlockouts
\usepackage{cite}
\usepackage{amsmath,amssymb,amsfonts}
\usepackage{graphicx}
\usepackage{mathrsfs}
\usepackage{float} 
\usepackage{array} 
\usepackage{textcomp}
\usepackage{xcolor}
\usepackage[linesnumbered,ruled,vlined]{algorithm2e} 
\usepackage{booktabs} 
\usepackage{multirow} 
\usepackage{threeparttable} 
\usepackage{makecell} 
\usepackage{subcaption} 
\usepackage{hyperref} 

\definecolor{customred}{HTML}{D52415}
\definecolor{customblue}{HTML}{424FCA}

\def\BibTeX{{\rm B\kern-.05em{\sc i\kern-.025em b}\kern-.08em
    T\kern-.1667em\lower.7ex\hbox{E}\kern-.125emX}}

\begin{document}
\title{KPG 193: A Synthetic Korean Power Grid\\Test System for Decarbonization Studies
\thanks{This work was supported by National Research Foundation of Korea (NRF) grant funded by the Korean government (MIST) (No. RS-2023-00210018) and KENTECH Research Grant  (202300008A).
}
}

\author{\IEEEauthorblockN{Geonho Song, Jip Kim}
\IEEEauthorblockA{\textit{Department of Energy Engineering} \\
\textit{Korea Institute of Energy Technology}\\
Naju, South Jeolla, Korea \\
\{geonhosong, jipkim\}@kentech.ac.kr}
}

\maketitle
\begin{abstract}
This paper introduces the 193-bus synthetic Korean power grid (KPG 193), developed using open data sources to address recent challenges of the Korean power system. 
The KPG 193 test system serves as a valuable platform for decarbonization research, capturing Korea's low renewable energy penetration, concentrated urban energy demand, and isolated grid structure. 
Clustering techniques were applied to preserve key system characteristics while maintaining computational tractability and representativeness. 
The system includes 193 buses, 122 generators, 407 transmission lines, and incorporates temporal weather datasets. 
Its feasibility was validated through Unit Commitment (UC) and AC Optimal Power Flow (ACOPF) simulations using 2022 demand and renewable generation data. 
This test system aims to provide a foundational framework for modeling and analyzing the Korean power grid.\end{abstract}

\section{Introduction}
As of 2022, the Asia-Pacific region accounts for over 50\% of global greenhouse gas emissions, with more than 80\% of the projected global coal demand growth originating from this region \cite{escap2022toward}. In Korea, coal contributes 33\% of the country's electricity supply, while renewables account for less than 10\% \cite{iea_korea}. Given the region's significant impact on global emissions, advancing open-source power grid modeling efforts in Korea is crucial to support decarbonization studies.
Europe has made substantial progress in open-source power grid modeling. The SciGrid \cite{SciGRID} and osmTGmod \cite{osmTGmod} projects modeled Germany's transmission network using OpenStreetMap (OSM), a publicly available dataset \cite{haklay2010good}. 
Building on this, PyPSA-Eur \cite{PyPSAEur} extended the scope to create an open model covering the entire ENTSO-E area. 
A key enabler in Europe is the public availability of geographical data for substations and transmission lines provided by ENTSO-E \cite{entsoe_map}. 
This openness allows for the validation of power grid models built on OSM data, enhancing their accuracy and reliability.
In the United States, open test systems have also achieved significant development. The pioneering work at Texas A\&M University \cite{TAMU} laid the foundation, later updated the system to meet 2016 standards with additional generators and high-voltage direct current (HVDC) transmission lines \cite{USTestSystem}. 
More recently, the TX-123BT network, which represents the Texas power grid and incorporates five years of meteorological data, has been made publicly available \cite{TX123BT}.

While regions such as Europe and the United States have made significant advancements in open-source power grid modeling through extensive research collaboration, synthetic power grid systems remain underdeveloped in Korea. 
Previous studies have utilized a realistic 1,428-bus system \cite{kim2022evaluation} and a reduced 11-zone system \cite{park2023clean}; however, only the 11-zone system is publicly available \cite{sendlab2023git}. 
To address this gap, we present the synthetic Korean power grid 193-bus test system (KPG 193), developed using open data from the year 2022. 
Detailed information about the Korean power grid is not publicly disclosed and publicly accessible data is limited to electricity demand and renewable energy capacity at the municipal level. 
To overcome these limitations while ensuring computational tractability and accurate grid representation, we developed the test system by clustering regions based on the locations of regional offices of the utility company. 
The proposed KPG 193 test system captures recent challenges of Korea's power grid, including concentrated electricity demand in metropolitan areas and the geographical distribution of power plants along the coastline. 
Moreover, the test system can be utilized for decarbonization studies and expansion planning research for Korea.
The main contributions of this paper are threefold:
\begin{itemize}
    \item \textbf{Development of the KPG 193 test system}: We provide synthetic transmission grid model for Korea based on open-source data, featuring 193 buses, 122 generators, and 407 transmission lines. The resulting KPG 193 is publicly available in \cite{KPGdata_egolab}.
    \item \textbf{Realistic modeling environment}: The KPG 193 test system simulates key characteristics of the Korean grid (e.g. power flow, voltage), serving as a testing environment for advanced power system modeling and analysis. 
    \item \textbf{Integration of renewables and weather datasets}: The system includes weather datasets used to generate renewable generation profile, enabling accurate simulations and facilitating advanced energy system studies.
\end{itemize}

\begin{figure*}[t]
    \centering
    \includegraphics[width=0.95\linewidth]{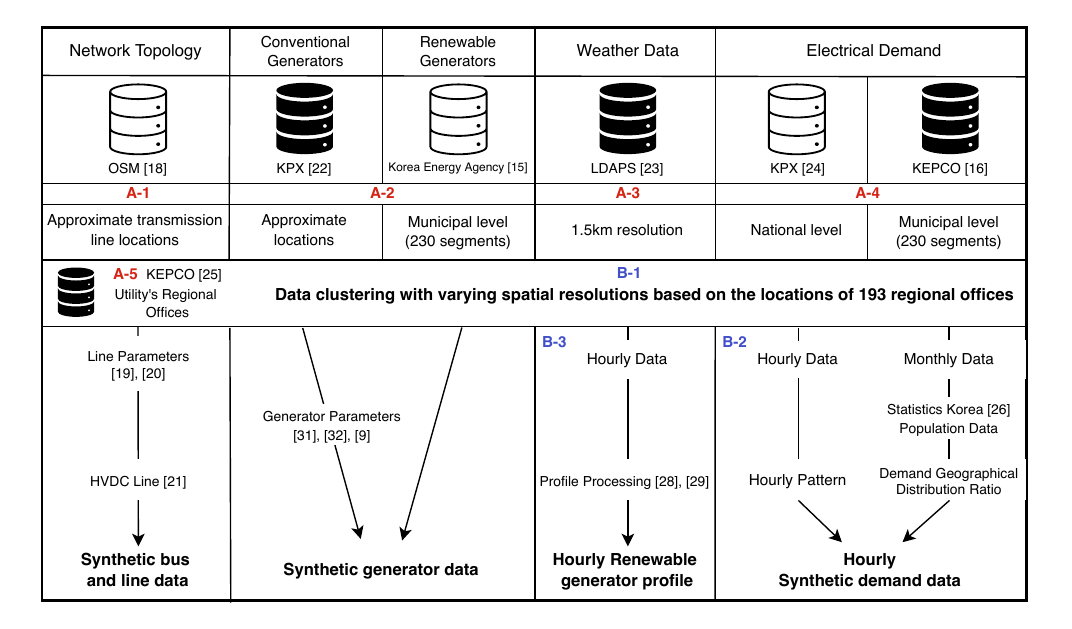}
    \vspace{-6mm}
    \caption{\small 
    Workflow for developing the KPG 193 test system with data sources and processes for modeling network topology, generator parameters, weather data, and electrical demand. \textcolor{customred}{Red} and \textcolor{customblue}{blue} letters correspond to \textit{subsections} in this paper that detail each process.
    }
    \label{fig:flowchart}
    \vspace{-4mm}
\end{figure*}

\section{Test System Modeling}
 
Developing a Korean test system using only open-source data presents two primary challenges. 
First, data on power plants and electrical loads is publicly available only at the municipal level \cite{generators_KoreaEnergyAgency, demand_KEPCO}, comprising 230 segments.
This limitation introduces considerable uncertainty in determining the locations of power plants and allocating electrical loads to specific buses. 
Second, there is no official source for network topology. 
Although open-source network data can be obtained from the OSM datasets \cite{haklay2010good}, it is known to have issues with topology accuracy as reported in reference \cite{openmscigrid2017}.

To address these challenges, we developed the test system following the workflow illustrated in Fig. \ref{fig:flowchart}. 
Datasets on network topology, generators, weather data, and electrical loads were initially collected to construct the power system. 
To ensure appropriate spatial resolutions and enhance accuracy, we applied clustering to network topology, generators, renewables, and demand data based on the locations of regional offices of the Korea Electric Power Corporation (KEPCO), the sole transmission and distribution utility in Korea.
Each process is detailed in the following subsections.

\subsection{Data Collection}
\subsubsection{Network Topology}
Two types of line data were extracted from the OSM database using the Overpass API: \textit{power line} for overhead lines and \textit{power cable} for underground cables \cite{topology_osm_overpass}. 
Each line element includes attributes such as voltage level and the number of parallel lines. 
AC line parameters were calculated using standard conductor coefficients from manufacturer catalogs, considering voltage level and conductor type \cite{grigsby2007electric, iljinelectric2023}. 
A summary of these line parameters is provided in  Table~\ref{tab:line_parameters}. 
The Korean power grid also features a HVDC transmission line connecting the \textit{Bukdangjin} and \textit{Godeok} substations with a capacity of 1,500 MVA \cite{topology_hvdc}. Since this HVDC line is not included in the OSM database, it was manually added during data preparation.

\begin{table}[t]
    \captionsetup{
        justification=centering,
        labelsep=period,
        font=footnotesize,
        textfont=sc
    }
    \caption{AC Line Parameters \cite{grigsby2007electric},\cite{iljinelectric2023}.}
    \vspace{-1mm}
    \centering
    \label{tab:line_parameters}
    \setlength\tabcolsep{2.5pt}
    \resizebox{\columnwidth}{!}{ 
        \begin{tabular}{c|c|c|c|c|c|c}
        \toprule
        \multicolumn{1}{c|}{ Line} & \multicolumn{1}{c|}{ Voltage} & \multicolumn{1}{c|}{ Conductor} & \multicolumn{1}{c|}{ $R$} & \multicolumn{1}{c|}{ $X$} & \multicolumn{1}{c|}{ $B$} & \multicolumn{1}{c}{ rateA} \\
        \multicolumn{1}{c|}{ Type} & \multicolumn{1}{c|}{ Level} & \multicolumn{1}{c|}{ Type} & \multicolumn{1}{c|}{ [$\Omega$/km]} & \multicolumn{1}{c|}{ [$\Omega$/km]} & \multicolumn{1}{c|}{ [nF/km]} & \multicolumn{1}{c}{ [MVA]} \\
        \midrule
        \scriptsize \multirow{3}{*}{\makecell{Overhead \\ line}} & \scriptsize 154 kV & \scriptsize ACSR 410 mm$^2\times2$ & \scriptsize 0.037 & \scriptsize 0.337 & \scriptsize 12.9 & \scriptsize 452 \\
        & \scriptsize 345 kV & \scriptsize ACSR 480 mm$^2\times2$ & \scriptsize 0.031 & \scriptsize 0.332 & \scriptsize 13.1 & \scriptsize 1,086 \\
        & \scriptsize 765 kV & \scriptsize ACSR 480 mm$^2\times6$ & \scriptsize 0.010 & \scriptsize 0.332 & \scriptsize 13.1 & \scriptsize 7,290 \\
        \midrule
        \scriptsize \multirow{2}{*}{\makecell{Underground \\ cable}} & \scriptsize 154 kV & \scriptsize XLPE 1200 SQ & \scriptsize 0.015 & \scriptsize 0.017 & \scriptsize 250 & \scriptsize 354 \\
                                            & \scriptsize 345 kV & \scriptsize XLPE 2000 SQ & \scriptsize 0.009 & \scriptsize 0.021 & \scriptsize 210 & \scriptsize 953 \\
        \bottomrule
        \end{tabular}
        
    }
    \vspace{2mm}
    \captionsetup{
        justification=centering,
        labelsep=period,
        font=footnotesize,
        textfont=sc
    }
       \caption{Generation Capacity and Share of the test system \cite{generators_KPX, generators_KoreaEnergyAgency}}
    \vspace{-0.35mm}
    \centering
    \label{tab:power_plants}
    \setlength\tabcolsep{5.5pt} 
    \resizebox{\columnwidth}{!}{ 
        \begin{tabular}{l|cccccc|c}
        \toprule
        \multicolumn{1}{l}{} & \multicolumn{1}{|c}{Coal} & \multicolumn{1}{c}{Gas} & \multicolumn{1}{c}{Nuclear} & \multicolumn{1}{c}{Solar} & \multicolumn{1}{c}{Hydro} & \multicolumn{1}{c}{Wind} & \multicolumn{1}{|c}{Total} \\
        \midrule
        \scriptsize \makecell{Capacity {[GW]}} & \scriptsize 38.13 & \scriptsize 41.20 & \scriptsize 24.65 & \scriptsize 23.75 & \scriptsize 7.20 & \scriptsize 1.65 & \scriptsize 136.57 \\
        \midrule
        \scriptsize \makecell{Capacity Share {[\%]}} & \scriptsize 27.9\% & \scriptsize 30.2\% & \scriptsize 18.0\% & \scriptsize 17.4\% & \scriptsize 5.3\% & \scriptsize 1.2\% & \scriptsize 100\% \\
        \bottomrule
        \end{tabular}
    }
    \vspace{-4.5mm}
\end{table}

\subsubsection{Conventional and Renewable Generators}
The system operator of Korean grid, Korea Power Exchange (KPX), provides the approximate locations and capacities for conventional power plants, including coal, gas, nuclear, and hydro sources as of March 2023 \cite{generators_KPX}. 
Data for renewable generators were sourced from the Korea Energy Agency, which reports cumulative installed capacities by energy source up to 2022 \cite{generators_KoreaEnergyAgency}. 
The total capacity mix of the KPG 193 test system is summarized in in Table~\ref{tab:power_plants}.

\renewcommand{\arraystretch}{1}

\subsubsection{Weather Data}

Weather data were sourced from the Korea Meteorological Administration's Local Data Assimilation and Prediction System (LDAPS), a numerical weather prediction model with a spatial resolution of 1.5 km \cite{KMA_LDAPS}. 
Using LDAPS, an hourly dataset for the year 2022 was generated, including temperature at 2 meters above ground level, wind speed at 93 meters, and solar irradiance.

\subsubsection{Electrical Load}

Electrical load data were obtained from two sources. 
The first source is the hourly total power demand for the entire system, provided by KPX \cite{demand_KPX}. 
The second source is the monthly electricity consumption at the municipal level, published by KEPCO \cite{demand_KEPCO}.

\subsubsection{Utility's Regional Offices and Population}

KEPCO provides the locations of its regional offices and their corresponding service areas \cite{regional_office_Kepco}. 
Offices with unclear service areas or those disconnected from the mainland were excluded, resulting in a final selection of 193 offices. 
Population data for each service area were aggregated using 2022 census statistics from Statistics Korea \cite{statisticskorea2023census}.

\subsection{Clustering and Data Processing}

\subsubsection{Clustering Method}

To model the Korean power grid while preserving its key characteristics and ensuring computational efficiency, clustering techniques were applied using the 193 regional offices of KEPCO with QGIS software \cite{QGIS_software}. 
This approach integrates datasets with varying spatial resolutions and enables power grid modeling without disclosing the exact locations of power facilities. 
In the KPG 193 network, the regional offices serve as buses ($b\in\mathcal{B}$) in the power grid. 
The endpoints of power lines and cables were clustered to the nearest bus, forming the lines ($l\in\mathcal{L}$) that connect the buses. 
To prevent isolated buses, some line connection points were manually adjusted. 
Power plants were assigned to the nearest buses, and their capacities were divided into standard generator units, denoted by ($g\in\mathcal{G}$).

\subsubsection{Electrical Demand and Weather Data Allocation}

Synthetic hourly demand data for each bus $b\in\mathcal{B}$ and time $t\in\mathcal{T}$ were generated in two steps. 
First, the total system demand was distributed across municipal regions based on monthly consumption data. 
Second, the municipal demand was allocated to buses within each service area using a demand geographical distribution ratio. 
For municipal regions served by multiple offices, demand was divided proportionally based on the population within detailed service boundaries. 
Weather data, including temperature at 2 meters above ground level $T_{bt}$, wind speed at 93 meters $v_{bt}$, and solar irradiance $\phi_{bt}$, were extracted for each bus $b$ at time $t$.

\subsubsection{Renewable Energy Profile Generation}
Renewable energy profiles were generated for each energy source using different methods. 
The hydropower profile $U_t$ was derived from historical generation data provided by KPX \cite{profile_KPX}, representing uniform hydroelectric production across all buses. 
Wind power profiles $U_{bt}^{\text{WT}}$ and solar power profiles $U_{bt}^{\text{PV}}$ were calculated for each bus $b$ and time $t$ using following equations \cite{atwa2009optimal}. 
        \vspace{-1mm}
\begin{align}
    \begin{split}
        &U_{bt}^{\text{WT}} =
        \begin{cases} 
            0 & \text{if }~~ 0\leq v_{bt}\leq v^{\text{CI}} ~~\text{or}~~ v^{\text{CO}}\leq v_{bt} \\ 
            \frac{v_{bt}-v^{\text{CI}}}{v^{\text{R}}-v^{\text{CI}}} & \text{if }~~ v^{\text{CI}}\leq v_{bt} \leq v^{\text{R}} \\
            1 & \text{if }~~ v^{\text{R}}\leq v_{bt} \leq v^{\text{CO}} \\
        \end{cases}
    \end{split} \\
    \begin{split}
        &U_{bt}^{\text{PV}}= FF^{\text{PV}}V^{\text{PV}}(T_{bt}, \phi_{bt})I^{\text{PV}}(T_{bt}, \phi_{bt})
    \end{split}
\end{align}
These calculations incorporate constants $v^{\text{CI}}$, $v^{\text{R}}$, $v^{\text{CO}}$ and $FF^{\text{PV}}$, which represent the cut-in wind speed, rated wind speed, cut-out wind speed, and fill factor of solar cell, respectively.
\subsection{KPG 193 test system data}
\begin{figure}[t]
    \centering
    \includegraphics[width=1\linewidth]{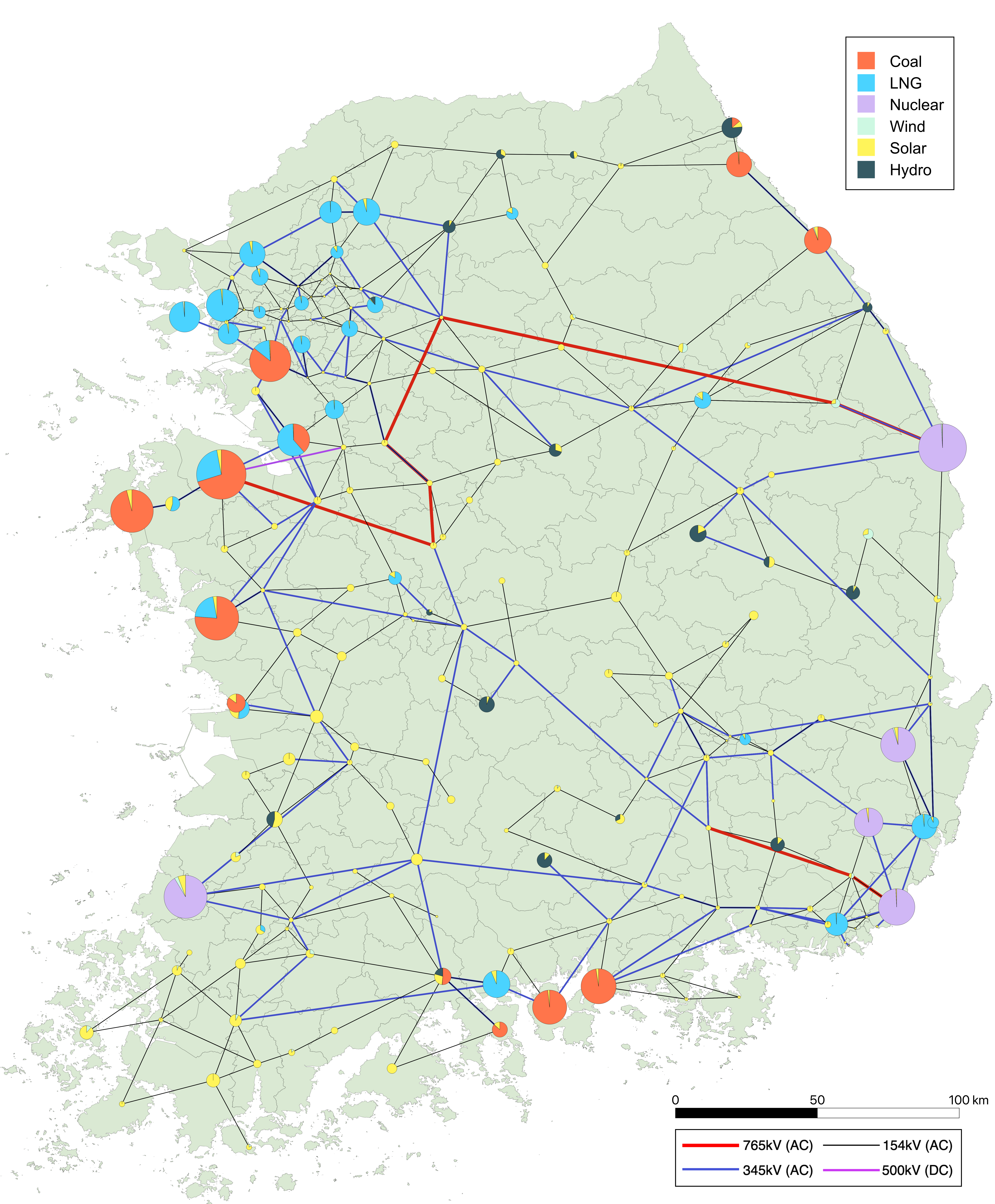}
    \vspace{-5mm}
    \caption{\small 
    Network topology and generation mix for the KPG 193 test system. Pie charts indicate the generation mix by fuel type, with sizes reflecting relative generation capacities. Transmission lines are distinguished by voltage levels, as shown in the legend.
    }
    \label{fig:system}
    \vspace{-2mm}
\end{figure}

The final KPG 193 test system is illustrated in Fig. \ref{fig:system}. 
The KPG 193 network comprises 193 buses, 122 generators, and 407 transmission lines across four voltage levels. 
Generator mix and capacity by energy source are visualized using a pie chart at each bus. 
The network data were formatted to comply with the MATPOWER standard \cite{matpower}, while bus-specific demand data, weather data, and renewable energy profiles over time were organized in CSV format. 
%

\begin{table}[t]
    \captionsetup{
        justification=centering,
        labelsep=period,
        font=footnotesize,
        textfont=sc
    }
    \centering
    \caption{Comparison of Annual Electric Generation Share by Fuel Type in 2022: Historical Data and KPG193}
    \vspace{-1mm}
    \label{table:generation_mix}
    \setlength\tabcolsep{5.5pt} 
    \resizebox{\columnwidth}{!}{%
        \begin{tabular}{l|ccccc|c}
            \toprule
            & Coal & Gas & Nuclear & Renewables & Etc.$^*$ & Total \\
            \midrule
            \scriptsize Historical {[\%]} & \scriptsize 32.5\% & \scriptsize 27.5\% & \scriptsize 29.6\% & \scriptsize 9.6\% & \scriptsize 0.8\% & \scriptsize 100\% \\
            \midrule
            \scriptsize KPG193 {[\%]} & \scriptsize 33.9\% & \scriptsize 27.8\% & \scriptsize 28.2\% & \scriptsize 10.1\% & \scriptsize 0\% & \scriptsize 100\% \\
            \bottomrule
        \end{tabular}%
    }
    \\[0.1mm]
    \begin{flushleft}
        \footnotesize $^*$ Etc. includes electricity generated from waste and oil.
    \end{flushleft}
    \vspace{-7.5mm}
\end{table}

\section{Validation}

\begin{table*}[t]
    \centering
    \renewcommand{\thetable}{\Roman{table}} 
    \captionsetup{
        justification=centering,
        labelsep=period,
        font=footnotesize,
        textfont=sc
    }
    \caption{Generator Parameters by Fuel Type: The parameters are adopted, randomized, and modified from \cite{lew2013western}, \cite{ISONewEngland} and \cite{USTestSystem}. The cost parameters ($C_g^{(2)}, C_g^{(1)}, C_g^{(0)}$) are presented as ranges across generator units for each fuel type.
    }
    \label{table:fuel_parameters_cost}
    \setlength\tabcolsep{5pt} 
    \renewcommand{\arraystretch}{1.2} 
    \begin{tabular}{l|ccccc|ccc}
        \toprule
        \multirow{2}{*}{Fuel Type} & {Min. Gen.} & {Ramp Rate} & {UT} & {DT} & {Startup Cost}  & {$C_g^{(2)}$} & {$C_g^{(1)}$} & {$C_g^{(0)}$} \\
        & [\% Cap.] & [\% Cap./hr] & [h] & [h] & [KRW/MW] & [KRW/MW$^2$h] & [KRW/MWh] & [KRW] \\
        \midrule
        LNG     & 52\%  & 100 & 4  & 3  & 53,862 &  2.1215 -- 6.6711  & 36,872 -- 70,956  & 637,657 -- 6,531,339 \\
        Coal    & 40\%  & 66  & 6  & 12 & 12,606 &  25.6102 -- 30.5675 & 22,912 -- 27,174  & 1,227,022 -- 2,629,634  \\
        Nuclear & 95\%  & 18  & 8  & 12 & -      &  1.6591 -- 3.0364 & 3,339 -- 8,292    & 0\\
        \bottomrule
    \end{tabular}
        \vspace{-2.25mm}
\end{table*}

The feasibility of the KPG 193 test system was evaluated by solving the daily Unit Commitment (UC) problem \cite{morales2013tight} and hourly AC Optimal Power Flow (ACOPF) simulations \cite{powermodels} for the entire year of 2022. 
Table \ref{table:generation_mix} presents the validation of generation results through comparison with actual system data \cite{profile_KPX}, with deviations within 1.4\%.
Since UC parameters and generation cost coefficients are not available, we derived parameters from \cite{lew2013western} and \cite{ISONewEngland}.
In addition, we modified the cost coefficients using methodologies from \cite{USTestSystem} to reduce discrepancies with historical data.
The quadratic generator cost function is:

\begin{figure}[t]
    \centering
    \vspace{-4mm}
    \includegraphics[width=1\linewidth]{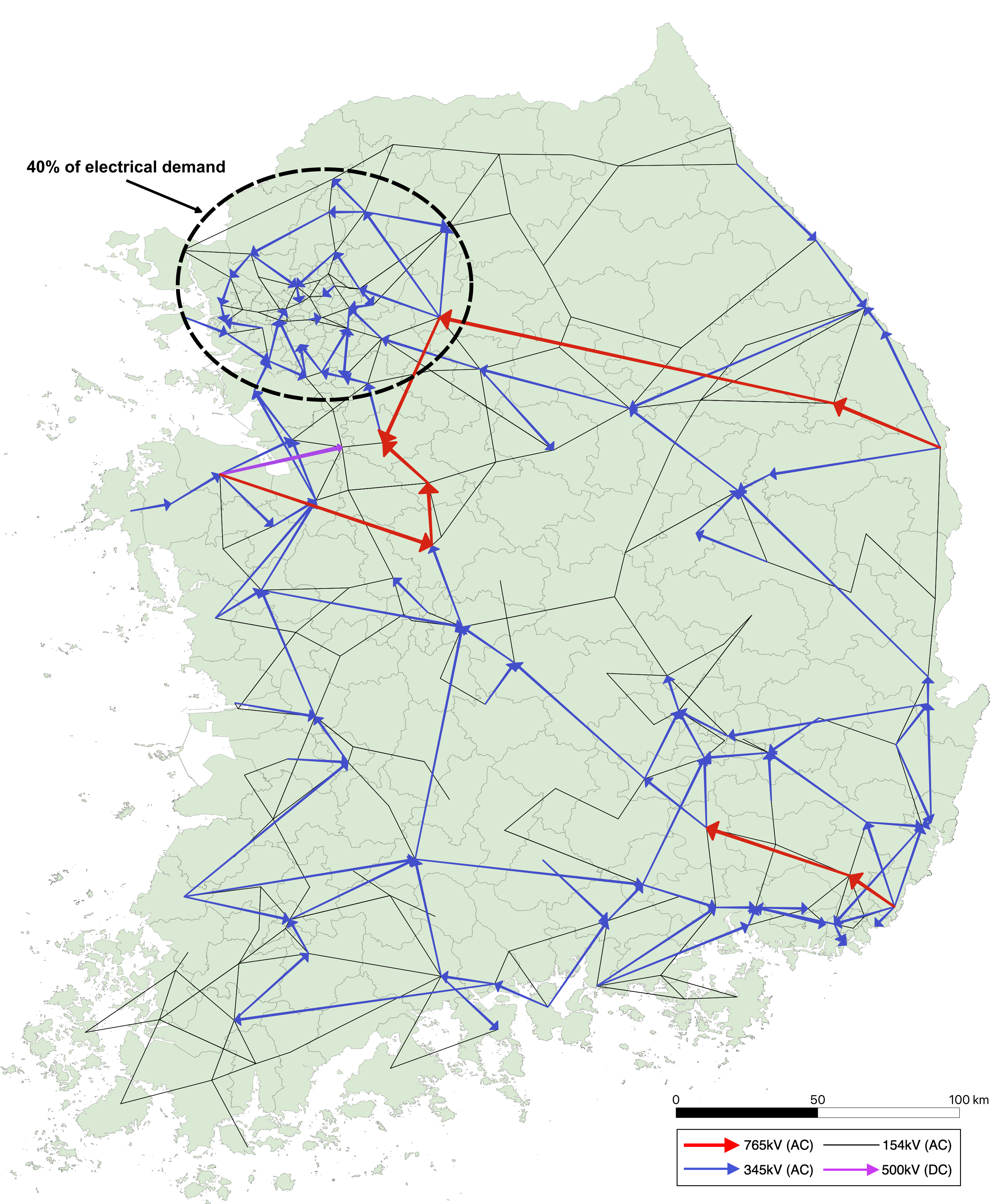}
    \vspace{-5mm}
    \caption{\small  Direction of power flow in 765kV and 345kV AC lines and 500kV DC line during the summer peak day at 13:00. The arrows indicate the direction of power flow, and the dashed circle highlights the region accounting for 40\% of the total electrical demand.}
    \label{fig:flow}
    \vspace{-5.6mm}
\end{figure}

\begin{figure}[t]
    \centering
    \vspace{-4mm}
    \includegraphics[width=1\linewidth]{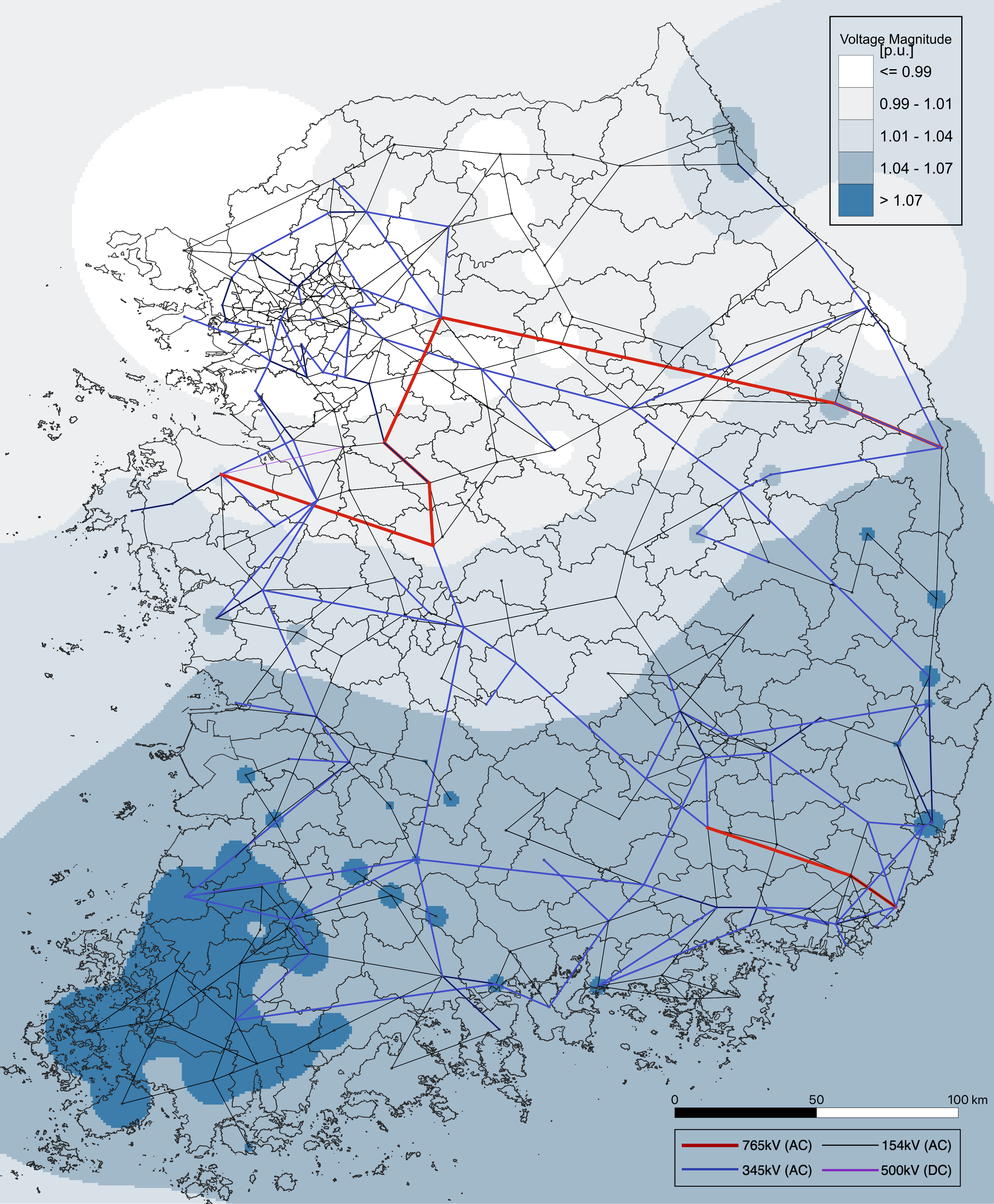}
    \vspace{-5mm}
    \caption{\small  Heatmap of nodal voltage magnitudes (p.u.) during the summer peak day at 13:00. The colors indicate voltage levels, with lighter shades representing lower voltages and darker shades representing higher voltages. Transmission lines are distinguished by voltage levels, as shown in the legend.}
    \label{fig:voltage3}
    \vspace{-6.5mm}
\end{figure}

\vspace{-5.5mm}
\begin{align}
        &c_{gt} = C^{(2)}_g p_{gt}^2 + C_g^{(1)} p_{gt} + C_g^{(0)}
\end{align}
where $p_{gt}$ is the power output of generator $g$ at time $t$, and $C^{(2)}_g, C^{(1)}_g$, and $C^{(0)}_g$ are the cost coefficients. Relevant parameters are detailed in Table \ref{table:fuel_parameters_cost}. 
When solving UC and ACOPF problems, nuclear generator maintenance data from 2022 is incorporated to reflect actual system conditions \cite{maintenance_KPX}.
Under these conditions, the daily UC and hourly ACOPF simulations converged across 2022 dataset.
The nuclear generator maintenance schedule in the KPG193 is available in \cite{KPGdata_egolab}.

Next, we validated whether the KPG 193 test system accurately reflects the key characteristics of the Korean power grid. 
The Korean power system is characterized by a concentrated of electricity demand in the metropolitan area, large-scale power plants along the eastern coast, and significant renewable energy integration in the southwestern region. 
These factors lead to predominant northward power flows toward the capital and overvoltage issues in regions with dense power plant clusters.

To verify these features, we conducted an ACOPF analysis using snapshot data from 13:00 on July 22, 2022, the peak demand day of the summer. 
As shown in Fig. \ref{fig:flow}, the circled metropolitan area accounts for approximately 40\% of the total power demand, while generators are concentrated along the coastal regions. 
This geographic distribution results in northward power flows through 345 kV, 765 kV, and HVDC transmission lines. 
In Fig. \ref{fig:voltage3}, regions with lighter colors indicate lower voltage magnitudes. 
Along the eastern coast, coal and nuclear power plants are densely clustered, transmitting power over distances exceeding 200 km to the metropolitan area. 
In contrast, the southwestern region with its high concentration of solar power installations, faces emerging overvoltage issues due to its relatively weak transmission network.
%
%

\section{Limitations}

The KPG 193 test system, developed exclusively using open data, inevitably faces several limitations due to the lack of detailed information about Korean power grid. Among these, the key limitations are as follows:
\begin{itemize}
    \item Various compensation devices, such as shunt capacitors, static synchronous compensators, and tap-changing transformers, are installed in the actual Korean power grid to mitigate voltage issues (illustrated in Fig.~\ref{fig:voltage3}). However, these devices were not included in the test system.
    \item Data sourced from OSM has uncertain update schedules, which may lead to inaccuracies in the power grid topology represented in the KPG 193 test system.
    \item Nodal reactive power demand is fixed at 90\% of the real power demand.
\end{itemize}

\section{Conclusion}

We present a methodology for devloping a synthetic test system using open-source data to advance power grid modeling in Korea. 
The data collection process included network topology, generator capacities, weather data, and electrical demand. 
The KPG 193 test system was created by clustering these datasets based on utility companies' regional offices, ensuring a computationally tractable model while preserving the key characteristics of the Korean power grid. 
The system organizes spatial and temporal data from 2022 to provide a comprehensive simulation environment.
The feasibility of the test system was validated by solving the daily UC and hourly ACOPF over an entire year. Additionally, a snapshot ACOPF analysis was performed to confirm that the KPG 193 test system replicates the characteristics of the Korean power grid. 
The simulation results successfully captured power flow and voltage patterns observed in the actual grid.


\begin{sloppypar}
\bibliographystyle{IEEEtran}
\bibliography{reference.bib}

\begin{thebibliography}{10}
\providecommand{\url}[1]{#1}
\csname url@samestyle\endcsname
\providecommand{\newblock}{\relax}
\providecommand{\bibinfo}[2]{#2}
\providecommand{\BIBentrySTDinterwordspacing}{\spaceskip=0pt\relax}
\providecommand{\BIBentryALTinterwordstretchfactor}{4}
\providecommand{\BIBentryALTinterwordspacing}{\spaceskip=\fontdimen2\font plus
\BIBentryALTinterwordstretchfactor\fontdimen3\font minus \fontdimen4\font\relax}
\providecommand{\BIBforeignlanguage}[2]{{%
\expandafter\ifx\csname l@#1\endcsname\relax
\typeout{** WARNING: IEEEtran.bst: No hyphenation pattern has been}%
\typeout{** loaded for the language `#1'. Using the pattern for}%
\typeout{** the default language instead.}%
\else
\language=\csname l@#1\endcsname
\fi
#2}}
\providecommand{\BIBdecl}{\relax}
\BIBdecl

\bibitem{escap2022toward}
{ESCAP, UN}, ``Toward sustainable energy connectivity in asia and the pacific: status, trends, and opportunities,'' 2022.

\bibitem{iea_korea}
\BIBentryALTinterwordspacing
{International Energy Agency (IEA)}, ``Korea - key energy statistics,'' 2024. [Online]. Available: \url{https://www.iea.org/countries/korea}
\BIBentrySTDinterwordspacing

\bibitem{SciGRID}
W.~Medjroubi and C.~Matke, ``Scigrid: Open source transmission network model, user guide (v 0.2),'' 2015.

\bibitem{osmTGmod}
M.~Scharf and A.~Nebel, \emph{osmTGmod 0.1.1: Documentation}, Wuppertal Institute, March 2016.

\bibitem{haklay2010good}
M.~Haklay, ``How good is volunteered geographical information? a comparative study of openstreetmap and ordnance survey datasets,'' \emph{Environment and Planning B: Planning and Design}, vol.~37, no.~4, pp. 682--703, 2010.

\bibitem{PyPSAEur}
J.~Hoersch \emph{et~al.}, ``Pypsa-eur: An open optimisation model of the european transmission system,'' \emph{Energy Strategy Reviews}, vol.~22, pp. 207--215, 2018.

\bibitem{entsoe_map}
\BIBentryALTinterwordspacing
{ENTSO-E}, ``{Transmission System Map},'' 2024. [Online]. Available: \url{https://www.entsoe.eu/data/map/}
\BIBentrySTDinterwordspacing

\bibitem{TAMU}
\BIBentryALTinterwordspacing
{Texas A\&M University}, ``Electric grid test case repository,'' 2024. [Online]. Available: \url{https://electricgrids.engr.tamu.edu/}
\BIBentrySTDinterwordspacing

\bibitem{USTestSystem}
Y.~Xu \emph{et~al.}, ``{US} test system with high spatial and temporal resolution for renewable integration studies,'' in \emph{2020 IEEE Power \& Energy Society General Meeting (PESGM)}.\hskip 1em plus 0.5em minus 0.4em\relax IEEE, 2020, pp. 1--5.

\bibitem{TX123BT}
J.~Lu \emph{et~al.}, ``A synthetic texas backbone power system with climate-dependent spatio-temporal correlated profiles,'' \emph{arXiv e-prints}, pp. arXiv--2302, 2023.

\bibitem{kim2022evaluation}
Y.-K. Kim \emph{et~al.}, ``Evaluation for maximum allowable capacity of renewable energy source considering ac system strength measures,'' \emph{IEEE Trans. on Sust. Energy}, 2022.

\bibitem{park2023clean}
W.~Y. Park \emph{et~al.}, ``A clean energy korea by 2035, transitioning to 80\% carbon-free electricity generation,'' 2023.

\bibitem{sendlab2023git}
\BIBentryALTinterwordspacing
SEND-LAB, ``South korea reduced network,'' 2023. [Online]. Available: \url{https://github.com/SEND-LAB}
\BIBentrySTDinterwordspacing

\bibitem{KPGdata_egolab}
\BIBentryALTinterwordspacing
``{Synthetic Korean Power Grid (KPG) 193-bus test system dataset},'' Nov 2024. [Online]. Available: \url{https://sites.google.com/view/ego-lab/resources/kpg-test-system}
\BIBentrySTDinterwordspacing

\bibitem{generators_KoreaEnergyAgency}
\BIBentryALTinterwordspacing
{Korea Energy Agency}, ``Supply statistics cumulative supply capacity - regional cumulative supply capacity (basic),'' July 2023. [Online]. Available: \url{https://www.knrec.or.kr/biz/statistics/supply/supply03\_03\_list.do}
\BIBentrySTDinterwordspacing

\bibitem{demand_KEPCO}
\BIBentryALTinterwordspacing
{KEPCO}, ``Electricity sales volume by city/county/district,'' February 2023. [Online]. Available: \url{https://home.kepco.co.kr/kepco/KO/ntcob/list.do?boardCd=BRD_000283&menuCd=FN05030105}
\BIBentrySTDinterwordspacing

\bibitem{openmscigrid2017}
Medjroubi \emph{et~al.}, ``Open data in power grid modelling: new approaches towards transparent grid models,'' \emph{Energy Reports}, vol.~3, pp. 14--21, 2017.

\bibitem{topology_osm_overpass}
\BIBentryALTinterwordspacing
M.~Raifer, ``{Overpass Turbo},'' 2024. [Online]. Available: \url{https://overpass-turbo.eu/}
\BIBentrySTDinterwordspacing

\bibitem{grigsby2007electric}
L.~L. Grigsby, \emph{Electric power generation, transmission, and distribution}.\hskip 1em plus 0.5em minus 0.4em\relax CRC press, 2007.

\bibitem{iljinelectric2023}
\BIBentryALTinterwordspacing
{ILJIN Electric}, ``\BIBforeignlanguage{Korean}{Catalog of extra-high voltage cable systems},'' 2023. [Online]. Available: \url{https://www.iljinelectric.co.kr}
\BIBentrySTDinterwordspacing

\bibitem{topology_hvdc}
\BIBentryALTinterwordspacing
{General Electric Company}, ``Transmitting power to cities: Ge's hvdc technology powers a densely populated south korean city,'' GE Grid Solutions, Tech. Rep., 2018. [Online]. Available: \url{https://www.gevernova.com/grid-solutions/products/applications/hvdc/bukdangjin-hvdc-lcc-casestudy-en-2018-04-grid-pea-0577.pdf}
\BIBentrySTDinterwordspacing

\bibitem{generators_KPX}
\BIBentryALTinterwordspacing
{KPX}, ``Transmission of electricity in south korea: Q\&a,'' July 2023. [Online]. Available: \url{https://www.kpx.or.kr/board.es?mid=a10504030000&bid=0048&act=view&list_no=69966}
\BIBentrySTDinterwordspacing

\bibitem{KMA_LDAPS}
\BIBentryALTinterwordspacing
{Korea Meteorological Administration}, ``Local data assimilation and prediction system (ldaps).'' [Online]. Available: \url{https://data.kma.go.kr}
\BIBentrySTDinterwordspacing

\bibitem{demand_KPX}
\BIBentryALTinterwordspacing
{KPX}, ``Hourly national electricity demand by city/county/district - 2022-12-31,'' February 2023. [Online]. Available: \url{https://www.data.go.kr/data/15065266/fileData.do\#layer\_data\_infomation}
\BIBentrySTDinterwordspacing

\bibitem{regional_office_Kepco}
\BIBentryALTinterwordspacing
{KEPCO}, ``Kepco introduction, organization, regional offices,'' November 2024. [Online]. Available: \url{https://home.kepco.co.kr/kepco/CU/CUIBPP002/main.do?menuCd=FN0101011002}
\BIBentrySTDinterwordspacing

\bibitem{statisticskorea2023census}
\BIBentryALTinterwordspacing
{Statistics Korea}, ``2023 population and housing census,'' 2023. [Online]. Available: \url{https://kosis.kr/eng/}
\BIBentrySTDinterwordspacing

\bibitem{QGIS_software}
\BIBentryALTinterwordspacing
{QGIS Development Team}, \emph{QGIS Geographic Information System}, QGIS Association, 2024. [Online]. Available: \url{https://www.qgis.org}
\BIBentrySTDinterwordspacing

\bibitem{profile_KPX}
\BIBentryALTinterwordspacing
{KPX}, ``2022 hourly generation capacity and power transaction volume by fuel type,'' March 2024. [Online]. Available: \url{https://epsis.kpx.or.kr/epsisnew/selectEkifBoardList.do?menuId=080100&boardId=010000}
\BIBentrySTDinterwordspacing

\bibitem{atwa2009optimal}
Y.~Atwa \emph{et~al.}, ``Optimal renewable resources mix for distribution system energy loss minimization,'' \emph{IEEE Transactions on Power Systems}, vol.~25, no.~1, pp. 360--370, 2009.

\bibitem{matpower}
R.~D. Zimmerman and C.~E. Murillo-Sanchez, \emph{MATPOWER User's Manual, Version 8.0}, MATPOWER, 2024.

\bibitem{lew2013western}
D.~Lew and G.~Brinkman, ``The western wind and solar integration study phase 2 (executive summary),'' National Renewable Energy Laboratory (NREL), Golden, CO (United States), Tech. Rep., 2013.

\bibitem{ISONewEngland}
D.~Krishnamurthy \emph{et~al.}, ``An 8-zone test system based on iso new england data: Development and application,'' \emph{IEEE Transactions on Power Systems}, vol.~31, no.~1, pp. 234--246, 2015.

\bibitem{morales2013tight}
Morales-Espa{\~n}a \emph{et~al.}, ``Tight and compact milp formulation for the thermal unit commitment problem,'' \emph{IEEE Transactions on Power Systems}, vol.~28, no.~4, pp. 4897--4908, 2013.

\bibitem{powermodels}
C.~Coffrin \emph{et~al.}, ``Powermodels. jl: An open-source framework for exploring power flow formulations,'' in \emph{2018 Power Systems Computation Conference (PSCC)}, 2018, pp. 1--8.

\bibitem{maintenance_KPX}
\BIBentryALTinterwordspacing
{KPX}, ``Weekly generator maintenance plan,'' 2022. [Online]. Available: \url{https://www.kpx.or.kr/board.es?mid=a10109030500\&bid=0018}
\BIBentrySTDinterwordspacing

\end{thebibliography}
\end{sloppypar}
\end{document}